\documentclass[
  aps,
  pra,
  reprint,
  superscriptaddress,
  amsmath,
  amssymb,
  longbibliography
]{revtex4-2}

\usepackage{graphicx}
\usepackage{bm}
\usepackage{hyperref}
\usepackage{xcolor}
\usepackage{soul}
\hypersetup{colorlinks=true,linkcolor=black,citecolor=black,urlcolor=black}

\newcommand{\Fkt}[1]{\,\mathsf{#1}}
\ifx\Tr\renewcommand{\Tr}{\Fkt{Tr}}\else\newcommand{\Tr}{\Fkt{Tr}}\fi

\begin{document}

\title{ The Quantum Polariton Hamiltonian that Reproduces the Same {\color{red}Mean Acceleration} High-Harmonic Generation Spectra as the Classical Hamiltonian in Strong Laser Fields}

\author{Nimrod Moiseyev}
\email{nimrod@technion.ac.il}
\homepage{https://nhqm.net.technion.ac.il/}
\affiliation{Schulich Faculty of Chemistry, Institute of Advanced Studies in Theoretical Chemistry, Faculty of Physics, and Solid State Institute, Technion--Israel Institute of Technology, Haifa 32000, Israel}

\begin{abstract}
This manuscript investigates the possibility of defining a quantum Hamiltonian that leads to the same {\color{red}mean-acceleration}  high-harmonic generation (HHG) spectra as those predicted by Floquet theory in the regime where the number of infrared (IR) pump photons is much larger than the number of emitted UV photons. The key assumption underlying our derivation is that the intensity of the emitted high harmonics is many orders of magnitude smaller than that of the IR laser, and that the emission of the $N^{nt}$ harmonic results from the absorption of $N$ IR photons. 
\end{abstract}
\maketitle

\section{Motivation}

{The fully quantum theory of high-harmonic generation $(HHG)$ by low-frequency laser fields was presented in {\color{red} Ref.\citenum{lewenstein1994theory},\citenum{corkum1993plasma}}, where an atom interacts with a continuous-wave (cw) classical electromagnetic field. The harmonic generation spectrum (HGS) is obtained from the mean acceleration, calculated by solving the time-dependent Schr\"odinger equation (TDSE), assuming that the atom is initially in its ground electronic state.
Instead of solving the Hermitian TDSE to compute the mean-acceleration HGS, one can employ non-Hermitian complex-scaling Floquet theory. Within this framework, the HGS is derived from the Floquet solution, which is dominated by the field-free ground state \cite{moiseyevWEINHOLDprl}.
{In this work, we focus on demonstrating that the mean-acceleration HGS obtained from Floquet theory is equivalent to the HGS derived from a quantum electrodynamics (QED) treatment based on a quantum polariton Hamiltonian, which is introduced here for the first time. The equivalence between the mean-acceleration HGS and the HGS obtained by solving the time-independent Schr\"odinger equation (TISE) of the polariton Hamiltonian holds under the following conditions:

(i) The initial complex-scaled Floquet solution, which yields the mean-acceleration HGS, is obtained from a Hamiltonian describing an atom interacting with infrared (IR) quantum light (i.e., QED IR fields).

(ii) The initial state,
$$
|\psi^{(0)}\rangle \otimes |0_{\mathrm{UV}}\rangle,
$$
corresponds to the vacuum of the ultraviolet (UV) QED Hamiltonian.

(iii) The final state,
$$
|\chi^{(0)}\rangle \otimes |1_{\mathrm{UV}}\rangle,
$$
is a degenerate eigenstate of the polariton Hamiltonian. This degeneracy arises when $N$ IR photons are transferred between Floquet channels, in the limit where the coupling between IR and UV radiation is neglected, and the UV mode is restricted to a single-photon excitation.

(iv) The coupling between the initial and final states is weak compared to the strength of the IR field and is linearly proportional to the symmetrized projector
$$
\frac{1}{\sqrt{2}} \left(
|\psi^{(0)}\rangle |0_{\mathrm{UV}}\rangle \langle \chi^{(0)}| \langle 1_{\mathrm{UV}}|
+
|\chi^{(0)}\rangle |1_{\mathrm{UV}}\rangle \langle \psi^{(0)}| \langle 0_{\mathrm{UV}}|
\right).
$$
This symmetric superposition of the two degenerate states emerges from a single-UV-photon interaction and ensures that the first-order correction to the polariton energy remains symmetric. {\color{red} The $1/\sqrt{2}$  factor is a phenomenological ansatz.} 

    The key assumption in our derivation is the absence of interference between the emitted UV photons and the absorbed IR photons. Any deviation from this condition leads to discrepancies between the HGS obtained from classical and quantum descriptions of the radiation field, even when the Floquet and QED IR Hamiltonians yield equivalent solutions. A numerical demonstration of this effect is presented.}

\section{Introduction}

 The quantum-optical nature of high-harmonic generation (HHG) has been studied for pulsed IR lasers. See for example\cite{FROMmatan-gorlach2023high,FROMmatanPhysRevA.61.043812}. In Ref.\citenum{OFERorenIDO-2020quantum}, it was shown that the harmonic generation spectrum (HGS) depends on the photon statistics. Since, under specific conditions \cite{moiseyev2024EVENTSUR}, the photon statistics can be obtained from Floquet calculations \cite{moiseyev2024photon}, one expects the HGS calculated using Floquet theory to coincide with the spectrum derived from QED calculations.
In Ref.\citenum{osti_21408513} Sindelka has shown that  the semiclassical approximation breaks down completely  in certain special but realistic cases, regardless of the fact that the incoming laser pulse contains a huge number of photons new effects arising due to the quantized nature of the radiation field. Here this special situation is not considered and 
the question we address here is how to define a quantum Hamiltonian in which the emission of UV photons results from the absorption of IR photons. Based on this Hamiltonian, we prove that the high-order harmonic generation (HHG) spectrum obtained from QED coincides with that derived from Floquet theory, provided that the laser field is sufficiently strong so that the number of infrared photons absorbed by the atom and subsequently emitted as high-frequency radiation is much smaller than the average number of photons in the laser mode.

We then briefly review the conditions under which the eigenfunctions of the QED Hamiltonian, dominated by the field-free atomic ground state, are approximately equal to the eigenfunctions of the Floquet Hamiltonian.
\section{Sketch of the proof}
\subsection{{\bf{FLOQUET} Hamiltonian for emitting one UV photon by absorbing $N_{HG}$ IR photons :}}

\begin{eqnarray}
 && \hat  H^{Floquet}= -i\hbar \frac{\partial}{\partial t}+\hat H_{atom} + \epsilon_{IR}^{Floquet}\hat d \cos(\omega_{IR} t)\nonumber ]] \nonumber \\ && - \epsilon_{UV} \frac{dV}{dx} \cos(\omega_{UV}t)
\end{eqnarray}
where $\omega_{UV}=N_{HG}\omega_{IR}$ and $V(x)$ is the effective one electron potential of the atom as a function of the electronic coordinate.\\

{\color{black}Why, in the interaction of an atom with a strong IR laser, do we use the length-gauge representation (i.e., the dipole interaction), whereas for the interaction with a weak UV laser we may use a different representation (regardless of whether the electromagnetic field is treated classically or quantum mechanically)?

The reason we use the dipole interaction for the strong IR laser is that, in either the Floquet or the QED formulation of the light–atom Hamiltonian, the Hamiltonian matrix becomes tri-diagonal. The dimension of this polaritonic Hamiltonian is given by the number of field-free atomic states used as a basis set multiplied by the large number of Floquet channels or QED photon states. This number must be large because the laser is very strong in order to generate high-harmonic generation (HHG).

By using the (t,t') method\cite{ttp}, we can work with a relatively small number of Floquet channels or QED states during the propagation of the initial state. The $(t,t')$ propagation method enables us to calculate the HHG spectrum even when the laser field is strong \cite{moiseyevWEINHOLDprl}.

Within the dipole approximation, the generation of high harmonics requires the calculation of the second-order time derivative of the time-dependent dipole moment, which yields the time-dependent acceleration. The Fourier transform of this time-dependent acceleration provides the HHG spectrum.

To avoid the need to calculate the time-dependent dipole moment explicitly, we use the acceleration gauge, in which the acceleration is linearly proportional to the first derivative of the atomic potential with respect to the light polarization direction\cite{moiseyev2022polariton, moiseyev2024photon}.}

Since $\epsilon_{UV}<< \epsilon_{IR}$ we use perturbation theory to calculate the HGS where,
\begin{eqnarray}
\label{H0_FLOQUET}
  &&  \hat H^{(0)}_{Floquet}=  -i\hbar \frac{\partial}{\partial t}+\hat H_{atom} + \epsilon_{IR}^{Floquet}\hat d \cos(\omega_{IR} t) \nonumber \\
  &&\hat  H^{(1)}=- \frac{dV}{dx} \cos(N_{HG}\omega_{IR}t)
\end{eqnarray}
Notice that we assume that the emitted UV photons do not interact one with another and actually we have set of zero order Hamiltonians for $N_{HG}=3,5,7,...$. 

\subsection{EIGENFUNCTION and SPECTRUM and HGS}
 The eigenfunction of the Floquet Hamiltonian that we will use to calculate the HGS is given by
\begin{equation}
|\Psi^{(0)}{Floquet}(x,t)\rangle= \Sigma_{n_f=-\infty}^{+\infty} |\phi^{Floquet}{n_f}(x)\rangle |n_f+\langle n_{photon}\rangle \rangle
\end{equation}
where $$\langle t|n_f+\langle n_{photon}\rangle\rangle=e^{i\omega_{IR}(n_f+\langle n_{photon}\rangle)t}$$.
This will be considered as the initial state, where the atom is in the vacuum state (no UV photon) of the UV radiation and in the presence of a strong IR laser with a mean photon number of $\langle n_{photon}\rangle$ (associated with $n_f=0$).

The $|\phi^{Floquet}{n_f}(x)\rangle$ are expanded in the basis set of the field-free eigenstates of the Hamiltonian,
\begin{equation}
|\phi^{Floquet}{n_f}(x)\rangle=\Sigma{n_{atom}}C^{Floquet}{n_{f},n_{atom}}|\varphi_{n_{atom}}(x)\rangle
\end{equation}
where $|\varphi_{n_{atom}}(x)\rangle$ are the field free solution of the TISE and  the selected coefficients $C^{Floquet}{n_{f},n_{atom}}$, which are the eigenvector components of the Floquet Hamiltonian matrix, are such that
\begin{equation}
C^{Floquet}{n_f,n_{atom}=1}>>C^{Floquet}{n_f,n_{atom}\ne 1}
\end{equation}

The corresponding zero-order Floquet energy is complex when complex scaling is used to calculate the photo-ionization rate of the open channel,
\begin{eqnarray}
E^{(0)}_{Floquet}= Er_{Floquet}^{(0)}-\frac{i}{2}\Gamma_{Floquet}^{(0)}
\end{eqnarray}

The time-independent mean acceleration amplitude is associated with the first-order correction to the Floquet energy, and the corresponding HGS are given by
\begin{eqnarray}
&& E^{(1)}_{floquet}(N_{HG})=\\ &&\langle \Psi^{(0)}_{floquet}(t)|-\frac{dV}{dx}\cos(N_{HG}\omega_{IR}t)| \Psi^{(0)}_{floquet}(t)\rangle_{space \&  time} \nonumber \\ 
&& = \int_{0}^{\infty }dt \Sigma_{n_f,n_f'} e^{-i\omega_{IR}(n_f-n'_f)t}\nonumber \\&&\langle \phi_{n_f}^{Floquet}|-\frac{dV}{dx}\cos(N_{HG}\omega_{IR}t)|\phi_{n_f'}^{Floquet}\rangle_x \nonumber \\ &&
=\Sigma_{n_f} \langle \phi_{n_f+N_{HG}}^{Floquet}|-\frac{dV}{dx}|\phi_{n_f}^{Floquet}\rangle
\end{eqnarray}
Therefore, 
\begin{equation}
    \sigma^{Floquet}(N_{HG})=\epsilon_{UV}^2\Big|E^{(1)}_{Floquet}\Big|^2
    \label{HGSfloquet}
\end{equation}

{\textit{Notice that Eq.\ref{HGSfloquet} holds  if and only if $\epsilon_{UV}<<\epsilon_{IR}^{Floquet}$ is equal to the HGS intensity as obtained by calculating the transform Fourier of the acceleration calculated by Floquet.}

\subsection{{\bf{\color{black}QED}: Hamiltonian for emitting one UV photon by absorbing $N_{HG}$ IR photons }}

\begin{eqnarray}
  &&  \hat H^{QED}=  \hat H_{atom} +\hbar \omega_{IR}\hat a^\dagger\hat a +\epsilon_{IR}^{QED}\hat d (\hat a^\dagger+\hat a)+\\ && \hbar \omega_{UV}\hat a^\dagger_{UV}\hat a_{UV}  \nonumber \\&& -\epsilon_{UV} \frac{dV}{dx} (\hat a_{UV}^\dagger+\hat a_{UV}) \frac{|\Psi^{(0)}_{QED}\rangle\langle\chi^{(0)}_{QED}|+|\chi^{(0)}_{QED}\rangle\langle\Psi^{(0)}_{QED}|}{\sqrt{2}}\nonumber 
   \label{QED-HAM}
\end{eqnarray}

where  we  assume that the driving is monochromatic and has a constant amplitude through out (and therefore excludes resonance phenomena due to the pulse rising and falling),  $\hat d$ is the dipole operator, and $\epsilon_{IR}^{QED}=\epsilon_{IR}^{Floquet}/{\sqrt{\langle  n_{photon}\rangle }}$ (see Ref.\citenum{moiseyev2023qed,li2022molecular,zhou2024nature}).
   The ratio between $\epsilon_{IR}^{Floquet}\equiv \epsilon^{QED}_{effective-IR}$ and emitted UV radiation $\epsilon_{UV}$ is about $5-9\cdot 10^{10}$\cite{REVIEW2006femtosecond}.
   {\color{red} The $1/\sqrt{2}$ factor is a phenomenological ansatz. This ansatz is motivated by the fact that the two Floquet resonances coupled by the cavity are narrow resonances, i.e., their complex poles lie close to the real energy axis. In the limiting case where these poles lie exactly on the real axis, tuning the cavity frequency to resonance would render the two Floquet states degenerate. The atom--cavity coupling would then lift this degeneracy, giving rise to upper and lower polaritons whose wavefunctions are equal-weight superpositions of the two Floquet states, with coefficients $\pm \frac{1}{\sqrt{2}}.$ Furthermore,, the same equal-weight superposition is expected in the case of a non-Hermitian degeneracy (an exceptional point, EP) that is obtained for specific values of the cavity parameters. At the EP, the atom--cavity interaction also lifts the degeneracy and generates upper and lower polaritonic states, whose wavefunctions are naturally described by the same linear combinations,$\pm \frac{1}{\sqrt{2}},$
of the two Floquet states.}
{\color{black} Persuasible due to rate of spontaneous emission which is proportional to the third power of the frequency. Therefore, the intensity of the emitted UV in HGS is much smaller than the intensity of the IR laser, and the interaction of the emitted UV photons with the atom is several orders of magnitude weaker than the interaction of the atom with the strong IR laser.

The initial function is the atom in the strong laser field with averaged number of IR photons ($\langle n_{IR}^{photon}\rangle$) and in the vacuum of the UV photons,  $|\Psi^{(0)}_{QED}\rangle$ and the final state $|\chi^{(0)}_{QED}\rangle$ is the atom in the slighter weaker laser  field with smaller averaged    number of photons which is  $\langle n_{IR}^{photon}\rangle-N_{HG}$ photons and in the field of one UV photon. Both $|\Psi^{(0)}_{QED}\rangle$ and $|\chi^{(0)}_{QED}\rangle$ are two degenrated  eigenfunctons of $$\hat H^{(0)}_{QED}= \hat H_{atom} +\hbar \omega_{IR}\hat a^\dagger\hat a +\epsilon_{IR}\hat d (\hat a^\dagger+\hat a)+ \hbar \omega_{UV}\hat a^\dagger_{UV}\hat a_{UV}  $$
Such that the initial state is given by,
\begin{eqnarray}
&& |\Psi^{(0)}_{QED}\rangle \\&&=|0_{UV}\rangle\Sigma_{n_{photon}=-\langle  n_{photon}\rangle}^{+\infty} |\phi^{QED}_{n_{photon}}(x)\rangle |n_{photon}+\langle  n_{photon}\rangle \rangle  \nonumber  
\end{eqnarray}
when for $n_{photon=0}$ the atom is in the field of IR laser with the initial $\langle  n_{photon}\rangle$ photons when the energy is equal to $\langle\Psi^{(0)}_{QED}| \hat H^{(0)}_{QED}|\Psi^{(0)}_{QED}\rangle=E_{IR}(N_{IR}=\langle  n_{photon}\rangle)$.
Whereas the final state is an atom in the weaker IR laser field with $\langle  n_{photon}\rangle-N_{HG}$ photons and a UV field of one photon is  given by,
\begin{eqnarray}
 &&   |\chi ^{(0)}_{QED}\rangle= \nonumber \\&& |1_{UV}\rangle \Sigma_{n_{photon}=-\langle  n_{photon}\rangle+N_{HG}}^{+\infty} |\varphi^{QED}_{n_{photon}-N_{HG}}(x)\rangle |n_{photon}\nonumber \\ &&+\langle  n_{photon}\rangle -N_{HG}\rangle
\end{eqnarray}

where, 
\begin{eqnarray}
&& |\varphi^{QED}_{n_{photon}}(x)\rangle=\Sigma_{n_{atom}}C^{QED}_{n_{photon},n_{atom}}|\varphi^{field-free}_{n_{atom}}\rangle \nonumber \\&&
C^{QED}_{n_{photon},n_{atom}=1}>>C^{QED}_{n_{photon},n_{atom}\ne 1}
\end{eqnarray}
When now $n_{photon=0}$ implies that the atom is in the field of IR laser with the initial $\langle  n_{photon}\rangle-N_{HG}$ photons while the energy is equal to $\langle\chi^{(0)}_{QED}| \hat H^{(0)}_{QED}|\chi^{(0)}_{QED}\rangle=E_{IR}(N_{IR}=\langle  n_{photon}\rangle-N_{HG})+\hbar\omega_{UV}$. For simplification of the derivation We now assume that $|\varphi^{field-free}_{n_{atom}}\rangle=|\phi^{field-free}_{n_{atom}}\rangle$. {\color{black}This is justified whenever the initial and final states are calculated from the zero order Hamiltonian where the coupling between then is taken as a very weak perturbation ($\epsilon_{UV}<< \epsilon_{IR}$)}\\
Since 
$$\omega_{UV}=N_{HG}\omega_{IR}$$ then  $\langle\chi^{(0)}_{QED}| \hat H^{(0)}_{QED}|\chi^{(0)}_{QED}\rangle= \langle \Psi^{(0)}_{QED}| \hat H^{(0)}_{QED}|\Psi^{(0)}_{QED}\rangle$.\\ 
Therefore, using the degenerate eigen-states $\Psi^{(0)}$ and  $\chi^{(0)}$ as two basis set the expectation value of  $\hat H^{QED}$ is given by,

\begin{eqnarray}
  &&  E^{QED} = E^{(0)}_{QED}+E^{(1)}_{QED}
  \end{eqnarray}
  where,

\begin{eqnarray}
E^{(0)}_{QED}= Er_{QED}^{(0)}-\frac{i}{2}\Gamma_{QED}^{(0)}  \\ 
 E^{(1)}_{QED}(N_{HG})=\nonumber \\ \frac{1}{2}\langle \chi^{(0)}_{QED}+ \Psi^{(0)}_{QED}|\hat  H^{(1)}_{QED}|\chi^{(0)}_{QED}+\Psi^{(0)}_{QED}\rangle \nonumber \\=\frac{1}{2}\langle \chi^{(0)}_{QED}|
\hat  H^{(1)}_{QED}|\chi^{(0)}_{QED}\rangle + \frac{1}{2}\langle \Psi^{(0)}_{QED}|\hat  H^{(1)}_{QED}|\Psi^{(0)}_{QED}\rangle \nonumber \\+\frac{1}{2}\langle \Psi^{(0)}_{QED}|\hat  H^{(1)}_{QED}|\chi^{(0)}_{QED}\rangle+\frac{1}{2}\langle \chi^{(0)}_{QED}|\hat  H^{(1)}_{QED}|\Psi^{(0)}_{QED}\rangle \nonumber \\  = 0+0 \nonumber \\
 +\frac{1}{2}\langle \Psi^{(0)}_{QED}|\hat  H^{(1)}_{QED}|\chi^{(0)}_{QED}\rangle  +\frac{1}{2}\langle \chi^{(0)}_{QED}|\hat  H^{(1)}_{QED}|\Psi^{(0)}_{QED}\rangle \nonumber \\
 = \Sigma_{n_{photon},n_{photon}'} \langle \phi_{n_{photon}+N_{HG}}^{QED}|-\frac{dV}{dx}|\phi_{n_{photon}'}^{QED}\rangle \nonumber 
\end{eqnarray}
and,

  \begin{eqnarray}
  &&\hat  H^{(1)}_{QED}=\\ &&- \frac{dV}{dx}  (\hat a_{UV}^\dagger+\hat a_{UV})\frac{|\Psi^{(0)}_{QED}\rangle\langle\chi^{(0)}_{QED}|+|\chi^{(0)}_{QED}\rangle\langle\Psi^{(0)}_{QED}|}{\sqrt{2}}\nonumber 
\end{eqnarray}

{\color{black}Degenerate  perturbation theory using  complex scaling theory and bi-orthogonal inner product rather than the standard Hermitian perturbation theory due to  associate a resonance state with a single square integrable function rather a wave packet in the continuum that is localized at the resonance energy with a width that approximately is the resonance width.}

\subsection{SPECTRUM and HGS}
Similar  to Eq.\ref{HGSfloquet} the HGS obtained from QED calculations of the energy correction due to the absorption of $N_{HG}$ IR photons that emit UV radiation with the frequency $\omega_{UV}=N_{HG}\omega_{IR}$ given by,
\begin{equation}
    \sigma^{QED}(N_{HG})=\epsilon_{UV}\Big|E^{(1)}_{QED}\Big|^2
\end{equation}
When $\phi^{QED}_{n_{photon}}=\phi^{Floquet}_{n_{f}}$ {\color{black}(whenever the conditions that the eigenvectors of the Floquet and QED Hamiltonoan matrix are non-distinguishable in the round off errors of the calculations as shown in  Ref.\citenum{moiseyev2024EVENTSUR} for the same model we solve here )} then \\
\begin{equation}
    \sigma^{QED}(N_{HG})=\sigma^{Floquet}(N_{HG})
\end{equation}

\subsection{{\color{black} Probability of creating a UV photon by losing 
$N_{HG} IR$ photons }}
Using the Lippmann-Schwinger expression for the transition probability one gets the same result as obtained above
\begin{eqnarray}
&& |\Psi^{initial}_{QED}\rangle= |\Psi_{atom}\rangle |n_{IR}^{photon}=\langle n_{IR}^{photon}\rangle\rangle|N^{UV}_{photon}=0\rangle \\ &&
 |\Psi^{final}_{QED}\rangle= |\Psi_{atom}\rangle |n_{IR}^{photon}=\langle n_{IR}^{photon}\rangle-N_{HG}\rangle|N^{UV}_{photon}=1\rangle\nonumber  \\ &&
T_{initial\to final}=\nonumber \\ && \Bigg| \frac{\langle \Psi^{initial}|\hat H^{(1)}{|\Psi^{(0)}_{QED}\rangle\langle}{\Psi^{(0)}_{QED}|}\hat H^{(1)}|\Psi^{final}\rangle}{E_{atom}+\hbar N_{HG}\omega_{IR}-E^{(0)}_{QED}}\Bigg|=\nonumber \\&&\frac{\sigma^{QED}(N_{HG})}{\Gamma^{(0)}_{QED}}=\frac{\sigma^{Floquet}(N_{HG})}{\Gamma^{(0)}_{Floquet}}
\end{eqnarray}
The last equality follows from the conditions proved in Ref. \citenum{moiseyev2024EVENTSUR}, where the spectra of the QED and Floquet Hamiltonians are almost identical.
\newpage

\section{ A Comparison of the Hamiltonians for Atomic Interaction with Classical and Quantum IR Electromagnetic Fields}
This comparison is based on Refs.\cite{moiseyev2024EVENTSUR,moiseyev2024photon} and is given for the coherency of the representation of our claim.
Under the dipole approximation, the time-dependent Hamiltonian of the system is given by Eq.\ref{H0_FLOQUET}
 To represent $ \hat H^{(0)}_{Floquet}$ in matrix form, we employ as a basis the time-periodic Fourier basis functions, $\{\exp(i n_f \omega_{IR} t)\}$, and the field-free eigenstates $\{|E_j^S\rangle\}_{j=1,2...,j_{max}}$ of the atomic system. With this basis, the matrix representation of the Floquet Hamiltonian is given by the {\it tri-diagonal matrix},
\begin{eqnarray}
\label{FLOQUET}
&& {\bf H_F}_{(j,n_f),(j',n_f')} =\\&&\big(E_j^S+\hbar\omega_{IR} n_f\big)\delta_{n_f',n_f}\delta_{j',j}+ \epsilon_{IR}^{Floquet}d^x_{j,j'}\delta_{n_f',n_f\pm 1}\nonumber
\end{eqnarray}
 in which $d^x_{j,j'}=\langle E^S_j|\hat d_x|E^S_{j'}\rangle$ is the dipole transition between two different atomic or molecular states along the laser polarization direction, \({n_f=0,\pm1,\pm 2 ,...,\pm (N_{Fourier}-1)/2}\), \(j,j'=1,...j_{max}\) and \(j_{max}\) is the dimension of the atomic Hilbert space. Practically, in numerical calculations, $N_{Fourier}$ and \(j_{max}\) are set to a finite value.
 \\Now, we assume that the same atom is placed in an optical cavity of volume \(V\). It is no longer driven by the laser field. Within the cavity, a single mode of the quantized electromagnetic field is described by the harmonic-oscillator {\it radiation} Hamiltonian (neglecting zero-point energy~\cite{milonni2013}),
$$\hat H_R=\hbar\omega_{IR}\hat a^\dagger \hat a \equiv \sum_{n=0}\hbar\omega_{IR} n |n\rangle \langle n| $$
Under the dipole approximation, the joint light-matter Hamitonian is given by 
\begin{eqnarray}
  \label{QED}
 {\bf H_{Q}}_{(j,n),(j',n')} &=& \Big(E_j^S+\hbar\omega_{IR} n\Big)\delta_{n',n}\delta_{j',j} \nonumber\\
 &+& e\epsilon_{IR}^{QED}d^x_{j,j'}\Big(\sqrt{n+1}\delta_{n',n+ 1}+\sqrt{n}\delta_{n',n- 1}\Big).\nonumber 
\end{eqnarray}
where the spectral representation of the atomic Hamiltonian was used $\hat H_A=\sum_j E_j^S|E_j\rangle\langle E_j|$. 
\textbf{Analytical equivalence of \(\bf H_F\) and \(\bf H_Q\) }.\\ We compare here between the spectrum of Floquet Hamiltonian matrix given in Eq.\ref{FLOQUET} and the spectrum of the QED Hamiltonian given in Eq.\ref{QED}. For the purpose of the comparison, we redefine the number of Floquet phtons $n_f$ in the Floquet Hamilotnian so that it takes the form of the following equation, 
\begin{eqnarray}
\label{FLOQUETmodified}
&& {\bf H_F}_{(j,n_f),(j',n'_f)} = \\ &&\Big(E_j^S+\hbar\omega_{IR} \big( \langle N_{ph} \rangle+n_f\big)\Big)\delta_{n'_f,n_f}\delta_{j',j}+ \nonumber \\ && \epsilon_{IR}^{QED}\sqrt{\langle N_{ph} \rangle}d^x_{j,j'}\delta_{n_f',n_f\pm 1}\nonumber 
\end{eqnarray}
where, 
\begin{equation}
   \epsilon_{IR}^{Floquet}=\sqrt{\langle N_{photon} \rangle}\epsilon_{IR}^{QED}
   \label{CEF}
\end{equation}
This amounts to adding a constant energy term to the Floquet Hamiltonian, i.e., the energy of the driving electromagnetic field. Notably, both the quantum and the semi-classical modified Floquet Hamiltonian matrices are tridiagonal block square matrices whose dimension is $j_{max}$ (i.e., the number of the field-free atomic or molecular eigenfunctions used as a basis set). To compare these Hamiltonians, we associate the number of Floquet channel to the number of QED photons by $$ 0\le n(Eq.\ref{QED})\,\equiv \,\langle N_{ph} \rangle+n_f \, (Eq.\ref{FLOQUETmodified}) $$ where $n_f=-\langle N_{photon} \rangle,..,0,\pm 1,\pm 2,..., +\langle N_{ph} \rangle$.

The comparison is done by subtracting the modified Floquet matrix from the QED Hamiltonian matrix, ${\bf {H_F}}-{\bf H_{Q}}$. 
 For sufficiently large value of $\langle N_{ph} \rangle$, all the ${\bf {H_F}}-{\bf H_{Q}}$ matrix elements approximately vanish since $\lim_{n\to \infty} \sqrt{( n \pm 1)/ n }=1$ . The size of this almost zero block matrix determines the number of absorbed or emitted photons that can be involved in multi-photo-induced dynamics of the atomic or molecular system under study.
 
 More rigorously,
using the Gershgorin circle (GC) theorem, the \textit{exact} eigenvalues of the quantum Hamiltonian given in Eq.\ref{QED} are bound inside a circle (in the complex plane), whose center is at ${\bf H_{Q}}_{(j,n),(j,n)} =E_j^S+\hbar\omega_{IR} n$, and its radius is 
$R_{GC} =\epsilon_{IR}^{QED}\sum_{j'}\big|d^x_{j,j'}\big(\sqrt{n+1}+\sqrt{n}\big)\big|$. When $n = \langle N_{photon} \rangle $, the center of the GC for the quantum Hamiltonian is the same as the GC for the modified Floquet matrix. 
When $\sqrt{n+1}+\sqrt{n}\approx 2\langle N_{photon} \rangle $ the radius of the GC for the modified Floquet is approximately equal to the radius of GC for the exact eigenvalues obtained for the quantum Hamiltonian. As $ n $ gets larger the eigenvalues and eigenvectors obtained from the QED calculations will be, to a good approximation, the eigenvalues and corresponding eigenvectors obtained from the Floquet calculations that are centered around a specific photon number $(\epsilon_{IR}^{Floquet}/\epsilon_{IR}^{QED})^2$. 
\section{Numerical study of a model Hamiltonian for a xenon atom in a laser field }

A model of Xe potential in
one spatial dimension with an atomic potential  has been
used before in the study of the high harmonic generation spectra in high-intensity laser fields \cite{fleischer2005NM}
\begin{equation}
  V(x)=-0.63\exp(-0.1424x^2)  
\end{equation}
where the parameters were fit to have two bound states, $E_0=-0.44 [a.u.]$ and $E_1=-0.14 [a.u]$ and a continuum.

The spectrum of this model Hamiltonian as obtained by rotating the continuum to the complex energy plane by $\theta= 0.3  $rad and diagonalizing the field-free Hamiltonian using
a particle in a box basis set, is presented in Fig.\ref{SPECTRUM}. 
\begin{figure}
\centering
\includegraphics[angle=000,scale=0.3]
{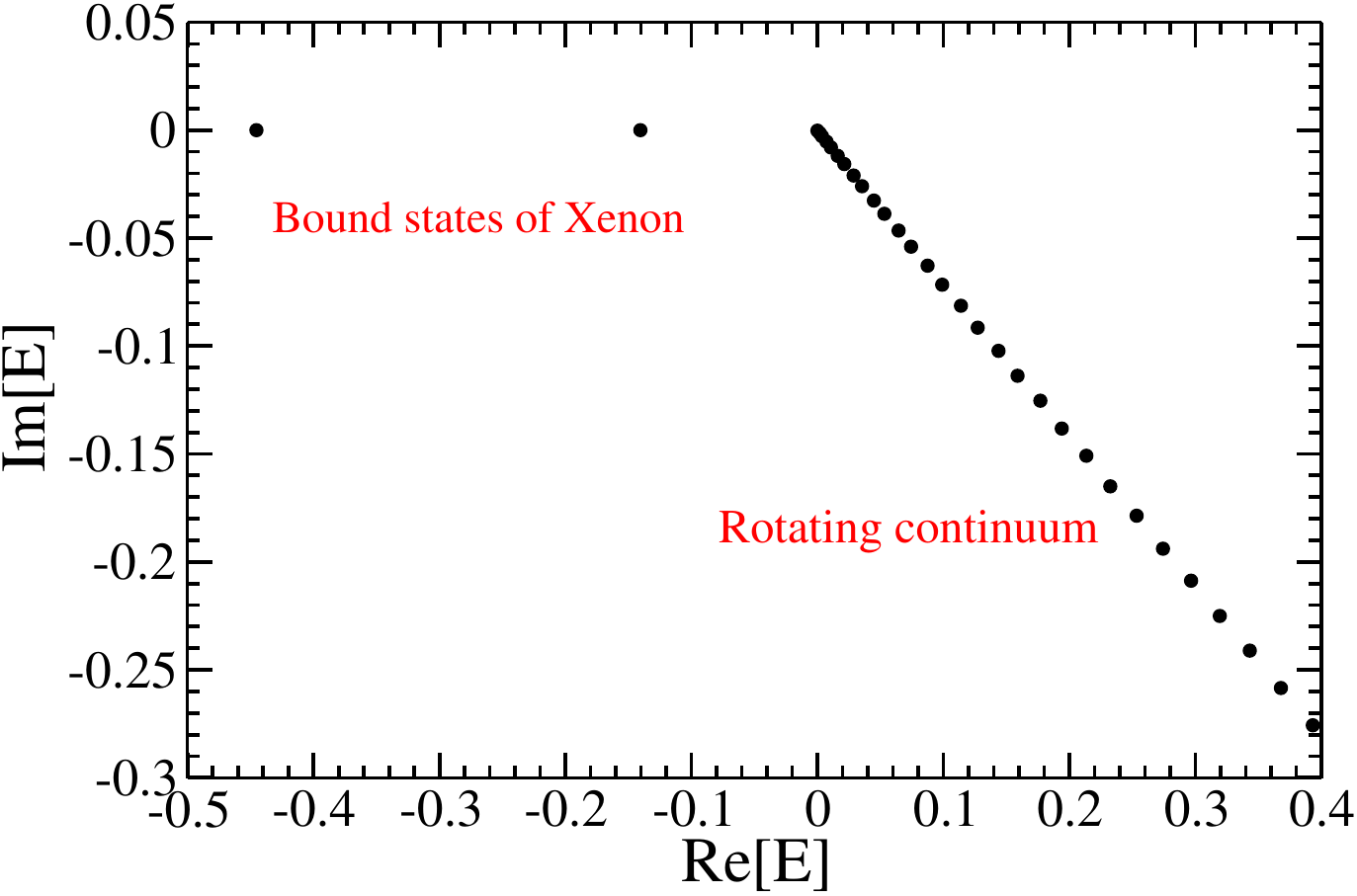}
\caption{The bound and rotating continuum as obtained from the diagonalization of the uniform complex scaled Hamiltonian matrix of Xe.}
\label{SPECTRUM}
\end{figure}
{\color{black}Since the ratio between the ionization energy and the IR photon energy is close to 11 it implies that for 500 absorbed IR photons the two bound states are embedded in the continuum of the atom that absorbed 489 photons and due to the coupling term in the Hamiltonian they become Feshbach type resonances associated with complex eigenvalues of the polariton Hamiltonian.
In Fig.~\ref{RATIO}, we plot the ratio between the square root of the number of absorbed IR photons, $\sqrt{n_{\mathrm{photon}}}$, as it appears in the QED Hamiltonian, and the square root of the average number of photons, $\sqrt{N_{\mathrm{photon}}}$, as it appears in the Floquet Hamiltonian. 
\begin{figure}
\centering
\includegraphics[angle=000,scale=0.3]{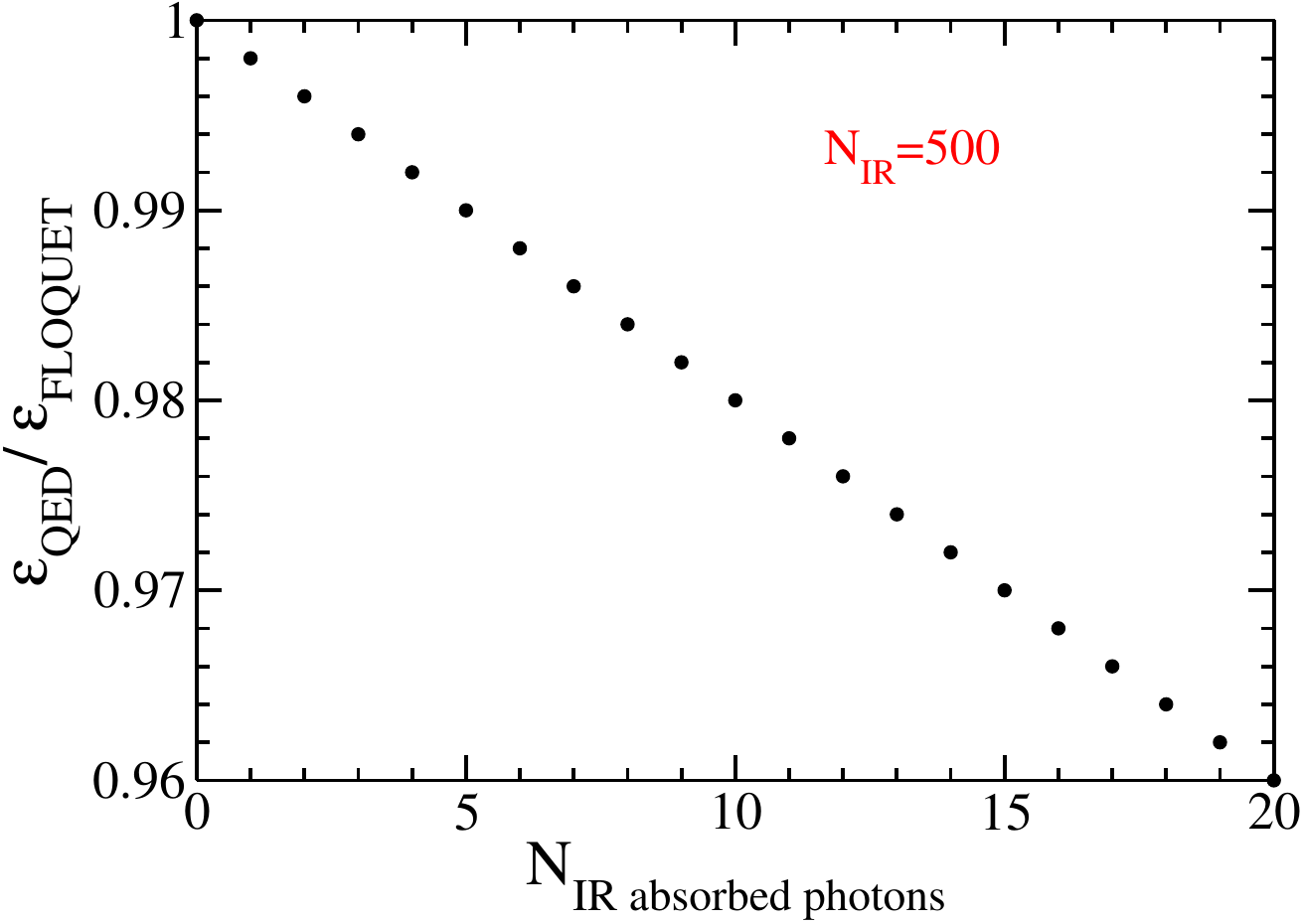}
\caption{Ratio between the square root of the number of absorbed IR photons, $\sqrt{n_{\mathrm{photon}}}$, as it appears in the QED Hamiltonian, and the square root of the average number of photons, $\sqrt{N_{\mathrm{photon}}}$, as it appears in the Floquet Hamiltonian. In red $N_{IR}=500$ 
 denotes the average number of photons in our model of a laser before the absorption of several IR photons. Here we show that even when $5$ IR photons are absorbed, the deviation between the QED and Floquet coupling strengths is about $1\%$. In realistic lasers, even when $100$ IR photons are absorbed, the QED and Floquet coupling strengths remain nearly identical. }
\label{RATIO}
\end{figure}
From the results presented in Fig.~\ref{RATIO}, we see that if the number of absorbed photons is less than 20, the QED and Floquet HGS are expected to be similar. We therefore calculated the eigenvectors of the QED and Floquet Hamiltonian matrices that have the largest overlap with the ground electronic state of the xenon atom and ensured that the dominant components are associated with a number of absorbed photons less than 20.
\begin{figure}
\centering
\includegraphics[angle=000,scale=0.3]{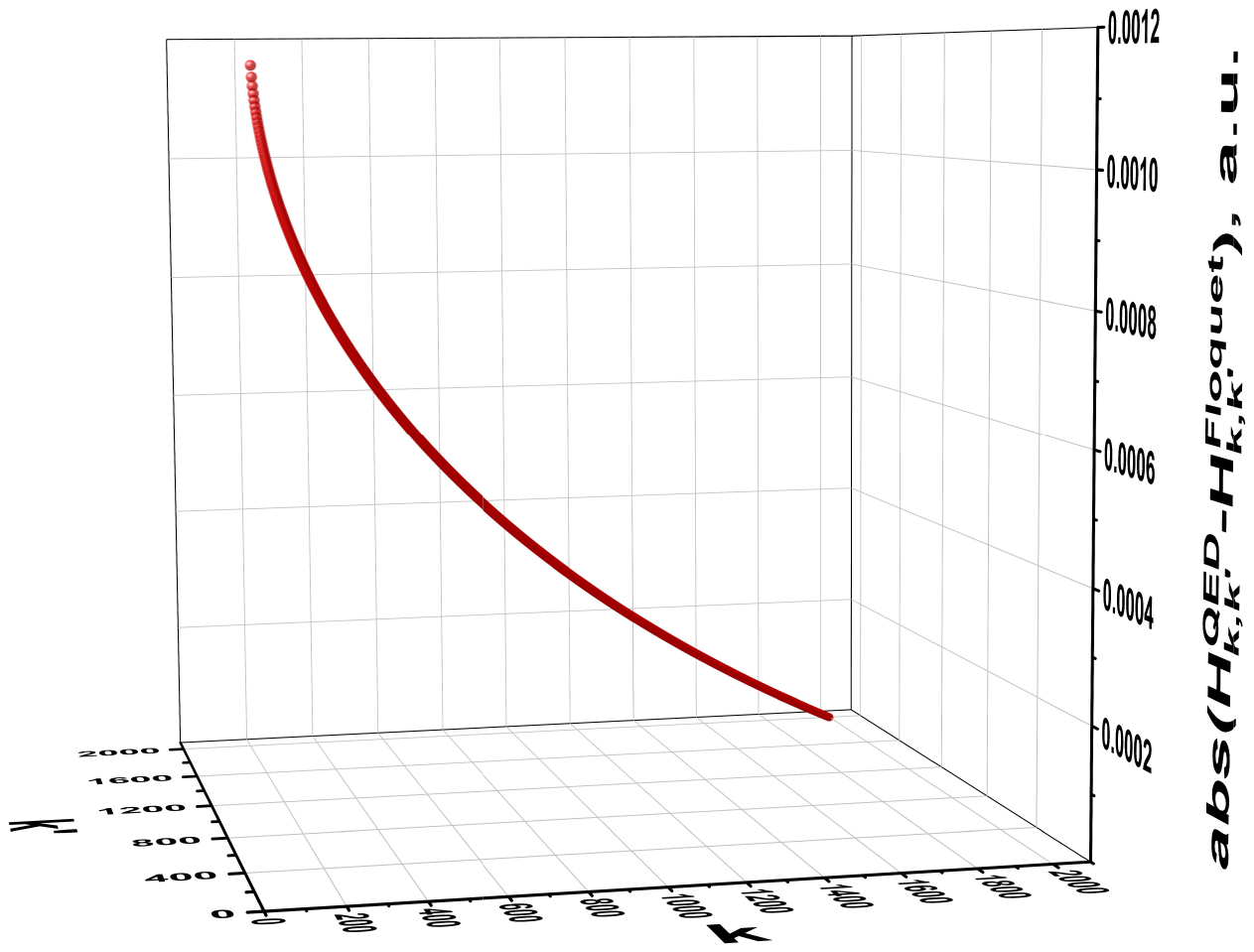}
\caption{The difference between the two Hamiltonian matrix elements of Xe, as obtained from interactions with classical and quantum electromagnetic fields.}
\label{DIFhamiltonianQED_FLOQUET}
\end{figure}
The difference between the QED and Floquet Hamiltonian matrix elements, shown in Fig.~\ref{DIFhamiltonianQED_FLOQUET}, is observed in the off-block-diagonal terms (the diagonal blocks are identical), which are very similar in both cases. In Fig.~\ref{RATIO}, we see that the blocks corresponding to 20 IR photons are nearly identical. Here we show that even more blocks of the two polaritonic Hamiltonians appear to be alike. The zoom presented in Fig.\ref{ZOOMDIFhamiltonianQED_FLOQUET} show how close are the off diagonal blocks in the two polaritonic Hamiltonians are alike. 

\begin{figure}
\centering
\includegraphics[angle=000,scale=0.3]{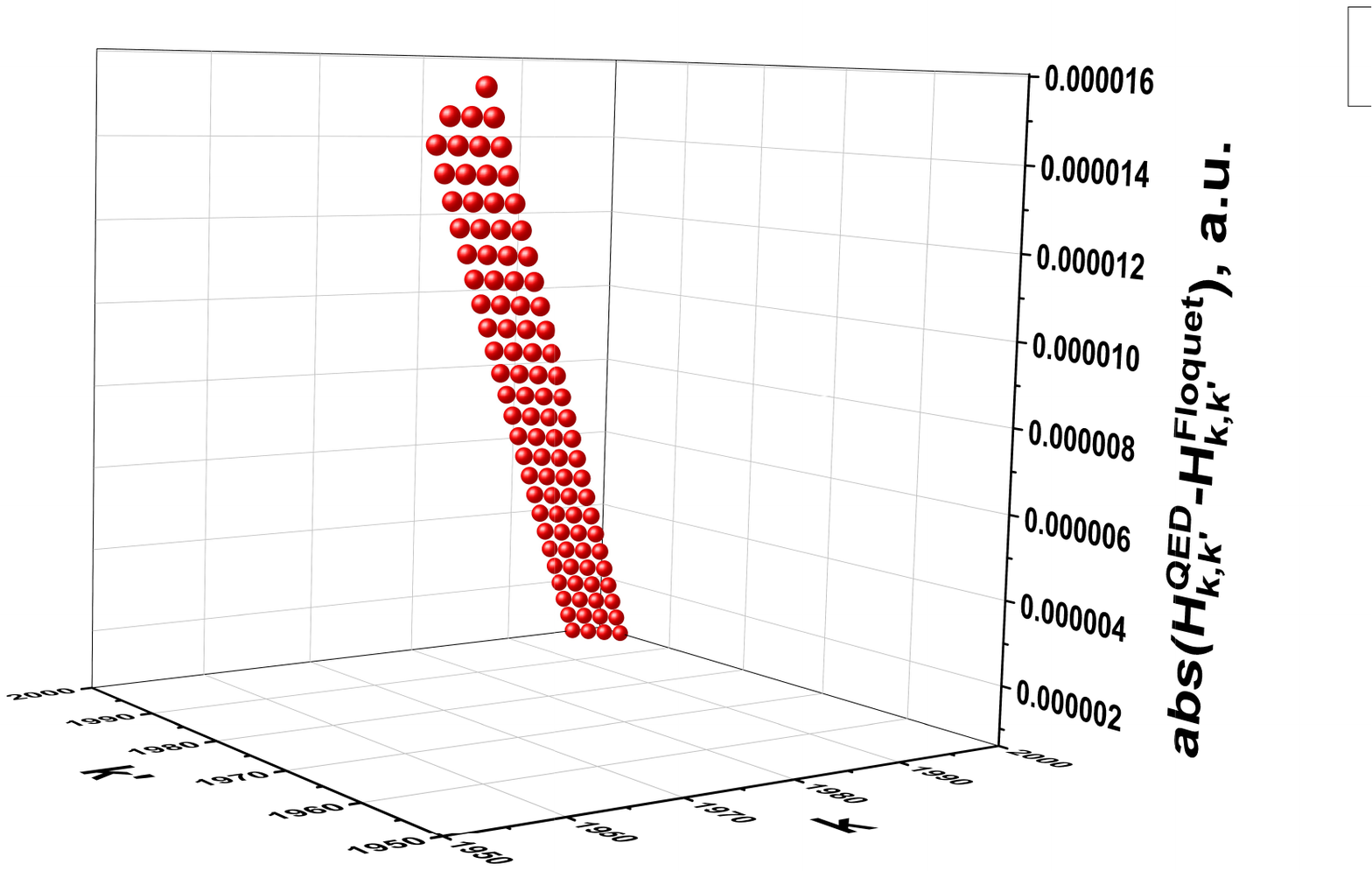}
\caption{Zoom on the difference between the two Hamiltonian matrix elements of Xe, as shown in Fig.\ref{DIFhamiltonianQED_FLOQUET}}
\label{ZOOMDIFhamiltonianQED_FLOQUET}
\end{figure}

In Fig.~\ref{ZOOMDIFhamiltonianQED_FLOQUET}, several off-diagonal matrix elements are nearly zero, regardless of the number of IR photons involved. Analysis of these elements shows that the QED and Floquet matrix elements are most similar when the electromagnetic field couples the ground state to the first excited bound state of xenon and the difference in number of absorbed photon is 1. For example if the $H^{QED,Floquet}_{k',k}=H^{QED,Floquet}_{k,k'}$ then  $k'=(1_{Xe},n_{photon}+1)$ and $k'=(2_{Xe},n_{photon})$ where the off diagonal matrix elements are degenerated since  $E_1^{Xe}+n_{photon}+1=E_2^{Xe}+n_{photon}$ and also with $k'=(1_{Xe},n_{photon})$ and $k'=(2_{Xe},n_{photon+1})$ where $E_1^{Xe}+n_{photon}\ne E_2^{Xe}+n_{photon+1}$. The first type of matrix elements are taken into consideration when we diagonalize the  Floquet or the QED Hamiltonian matrices.


Since the HGS is calculated from the eigenvectors of the QED and Floquet Hamiltonian matrices, it follows from the discussion above that these eigenvectors are expected to be nearly identical. Indeed, the results of the diagonalization of these two polaritonic Hamiltonians, presented in Fig.~\ref{Floquet_VS_QED}, confirm this expectation. 

\begin{figure}
\centering
\includegraphics[angle=000,scale=0.3]{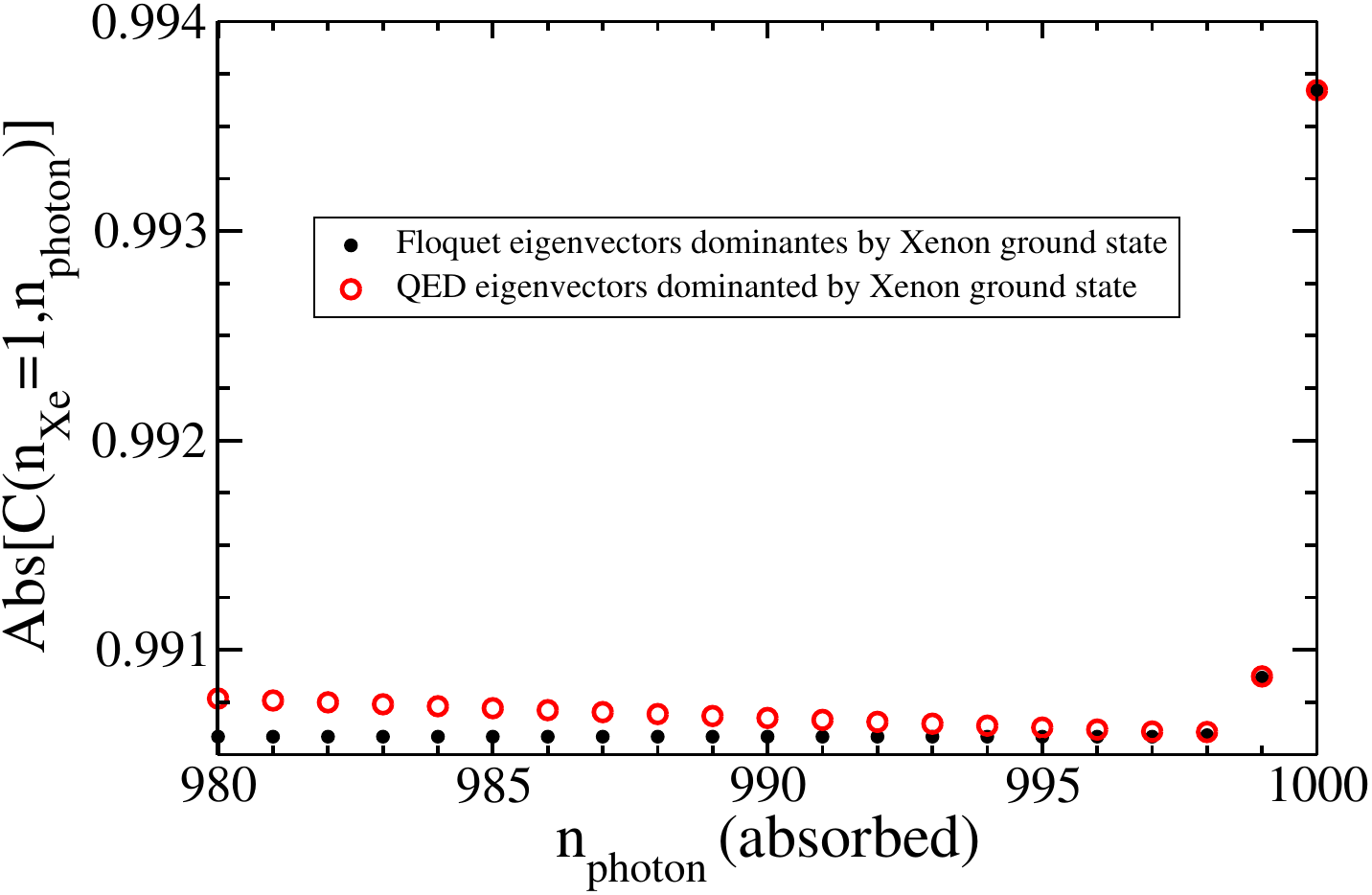}
\caption{A comparison of the eigenvectors of the Floquet and QED polariton Hamiltonian matrices dominated by the ground electronic state, as shown in Fig. \ref{SPECTRUM}.}
\label{Floquet_VS_QED}
\end{figure}

In the following plots we show the HGS as obtained from the diagonalization of the Floquet and QED polariton matrices for different number of IR photons that are observed by the Xenon in its ground state.

Due to the symmetry of Floquet and QED Hamiltonians $N_{HG}$ should get odd values only since the acceleration operator is an odd function.
\begin{eqnarray}
&&\Big|A(N_{HG})\Big|^2=\Big|\Sigma_{n_{photon}=0}^{{\color{red}N_{photons}}-N_{HG}}\Sigma_{n_{Xe}} \Sigma_{n'_{Xe}}\\&&C(n_{Xe},n_{photon}+N_{HG})Ac_{n_{xe},n'_{xe}}C(n'_{Xe},n_{photon})\Big|^2\nonumber
\label{HGS}
\end{eqnarray}
where $N_{photons}=1000$, $N_{HG}\le 100$ and ${\bf Ac}$ is the presentation of the acceleration operator $-\frac{dV}{dx}$ by a matrix, within the field-free resonance basis set. See Fig. \ref{HGS-Floquet_VS_QED} for a comparison of the HGS obtained using the Floquet method and the QED polariton matrix, where two bound states of xenon are used as the basis set with 1000 photon/Floquet channels.
\begin{figure}
\centering
\includegraphics[angle=000,scale=0.3]{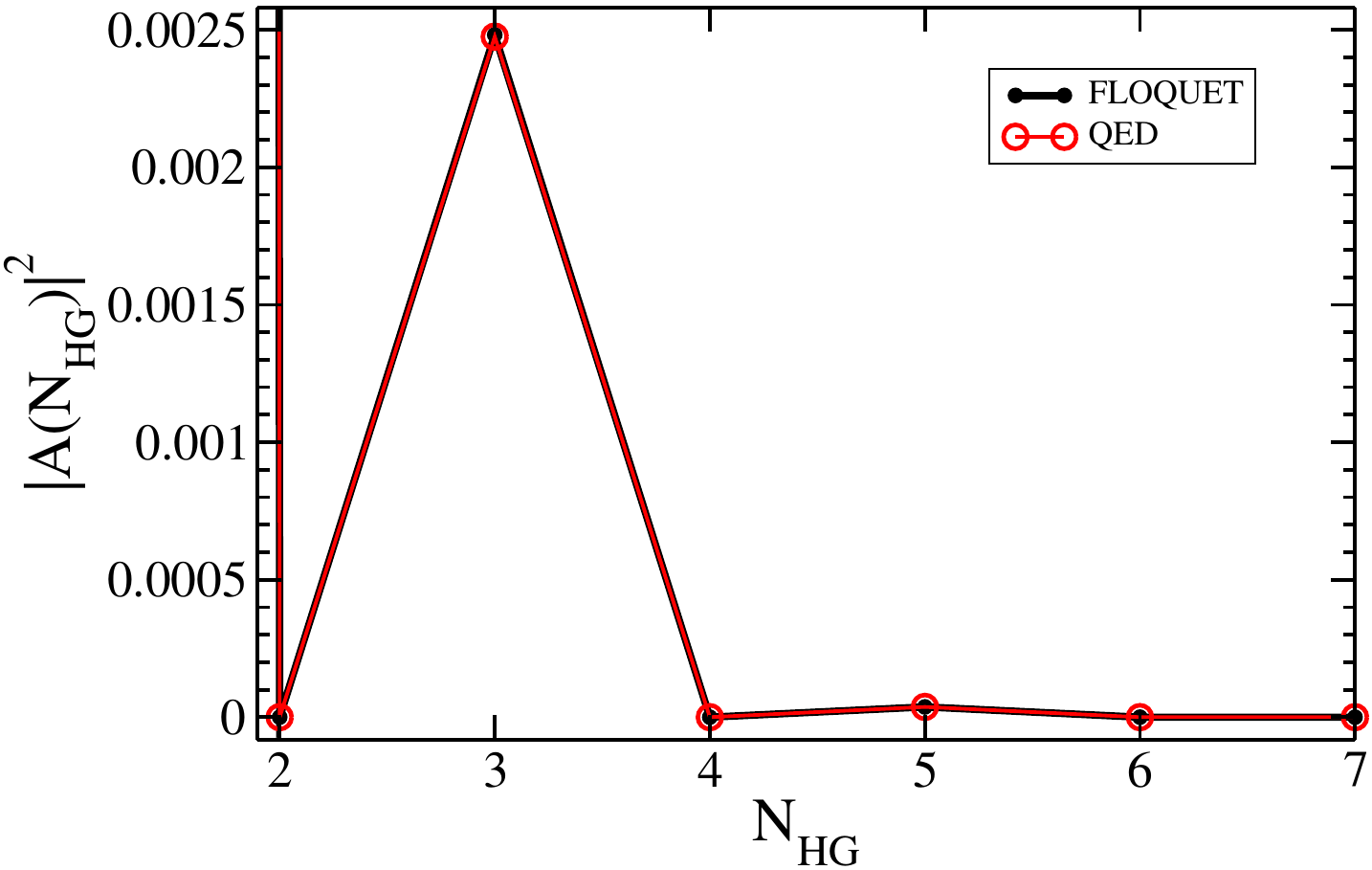}
\caption{A comparison of the HGS as obtained from the Floquet and QED polariton Hamiltonian matrices dominated by the ground electronic state, as shown in Fig. \ref{SPECTRUM}. The Floquet and QED HGS amplitudes where scaled to get $A(N_{HG}=1)=1$. }
\label{HGS-Floquet_VS_QED}
\end{figure}
{\it The excellent agreement between the HGS obtained from QED and Floquet  calculations supports the definition of the quantum polariton Hamiltonian as given in Eq.\ref{QED-HAM}}.
{\color{red}To show the convergence of the HGS with respect to the number of photons obtained from QED calculations, we used Eq.\ \ref{HGS}, where $N_{\text{photons}}$ is increased up to $1000\ \text{photons/cm}^3$ to calculate the intensity of the fifth high harmonic. Using Pade' approximation (i.e.,$ \sigma_{\Omega_{\text{HG}}=5\omega_{\text{IR}}}(N_{\text{photons}})$ is approximated as a ratio of two polynomials\cite{schlessingerPADE}), we extrapolate $\sigma_{\Omega_{\text{HG}}=5\omega_{\text{IR}}}(N_{\text{photons}})$ to $N_{\text{photons}}\to\infty$. The results of the extrapolation are shown in Fig.\ \ref{HGS-PADE}.}
\begin{figure}
\centering
\includegraphics[angle=000,scale=0.3]{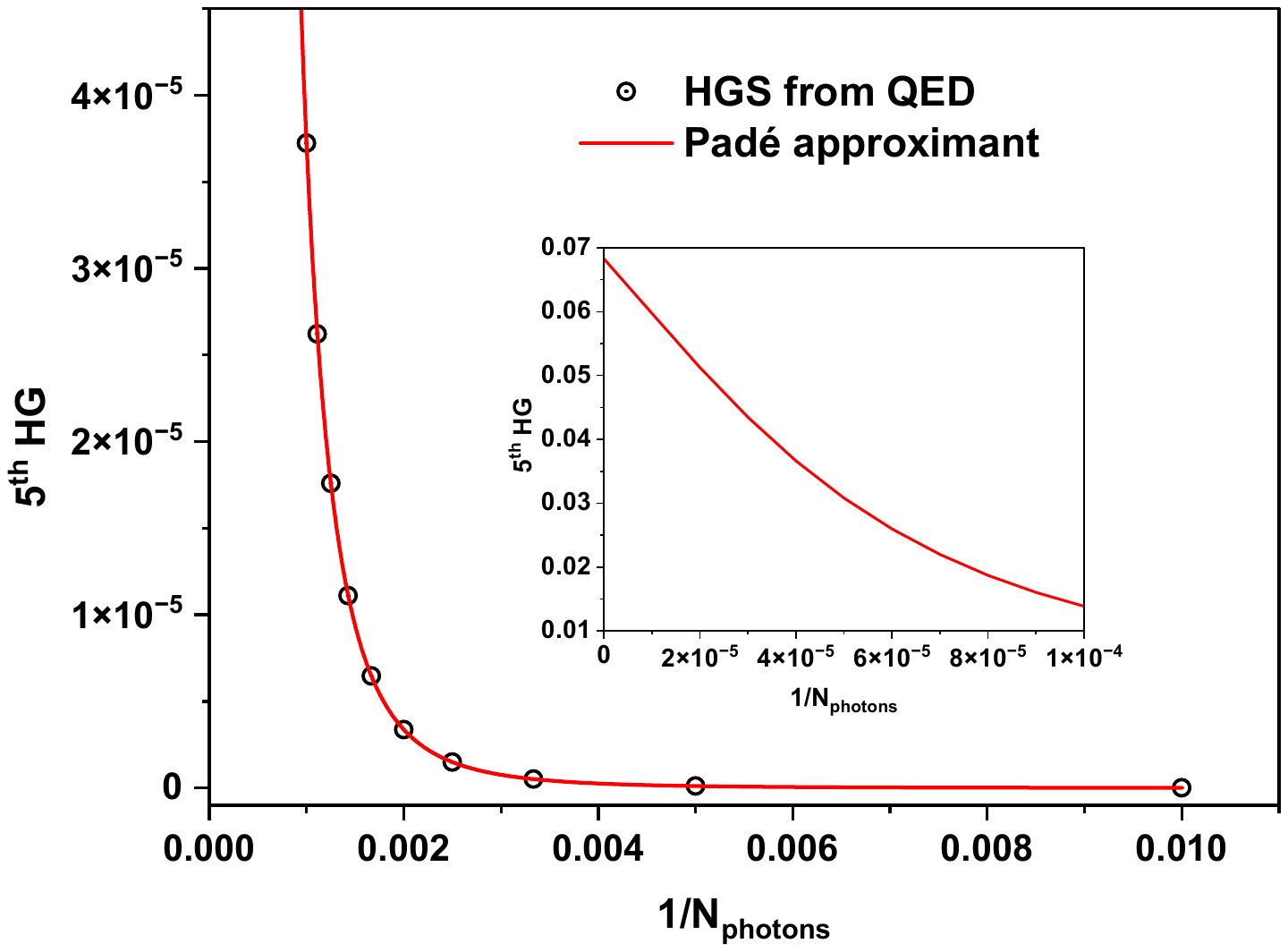}
\caption{{\color{red} The extrapolation of  $\sigma_{\Omega_{\text{HG}}=5\omega_{\text{IR}}}(N_{\text{photons}})$ to $N_{photons}\to \infty $ by Pade' approximate\cite{schlessingerPADE} is introduced. }}
\label{HGS-PADE}
\end{figure}
Even for strong laser fields, where the number of photons (equivalently, the number of Floquet channels) is extremely large, the Floquet eigenstates that are adiabatically connected to the field-free atomic states remain identical to the field-free states, up to numerical round-off errors. Consequently, the high-harmonic generation (HHG) spectrum obtained from the Fourier transform of the expectation value of the acceleration must be identical.

Below, we present the HHG spectrum on a logarithmic scale, calculated using the $(t,t')$ Floquet formalism with $2,!555,!000$ Fourier basis functions. The spectrum clearly exhibits both the characteristic plateau and the cutoff region of high-harmonic generation.
\begin{figure}
\centering
\includegraphics[angle=000,scale=0.3]{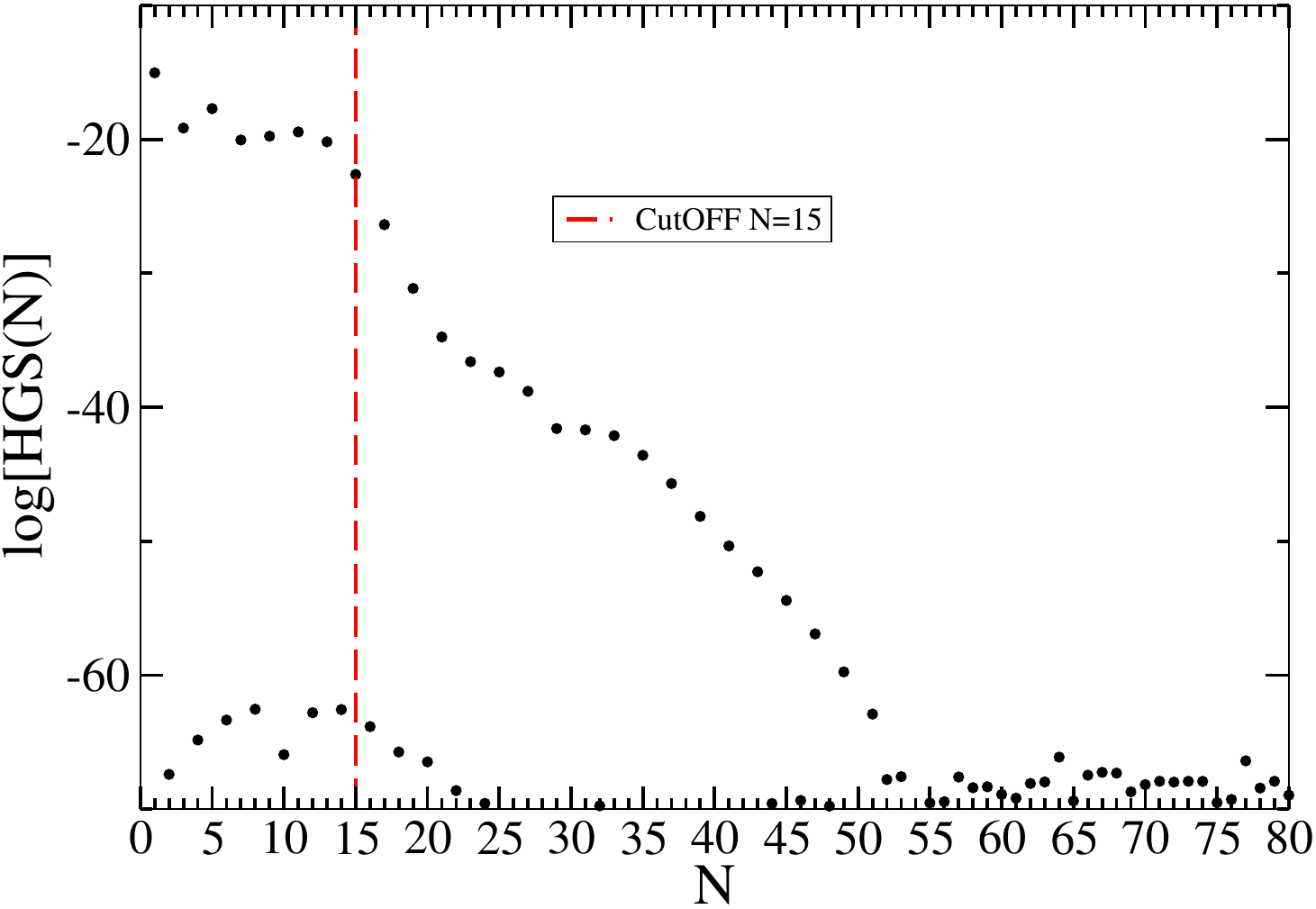}
\caption{HGS as obtained by ttp method for $\epsilon_0=0.03$ and $\theta=0.1$. The $Cutoff=[I.P + 3.17\epsilon_0^2/(4\omega^2)]/\omega$}
\label{HGSTTP}
\end{figure}
As can be seen from the results presented in Fig.~\ref{HGSTTP}, the number of significant high harmonics is small (approximately 15) compared with the number of IR photons in the driving laser field. To complete the justification of the proof based on Eq.~\ref{QED-HAM}, we calculate the ratio between the intensity of the high harmonics in the plateau region and the intensity of the first harmonic, corresponding to $\omega_{UV}=\omega_{IR}$. This ratio is found to be nearly zero, thereby justifying the use of the zero-order approximation in constructing the coupling between the atom and the quantized UV field. Our calculations, presented in Fig.~\ref{HGSTTP}, show that this ratio is smaller than $10^{-17}$.
\section{Concluding Remarks}
{\color{black}Here we show that when the classical field parameters (intensity, frequency, and photon statistics) used in the QED calculations are the same as those used in the Floquet calculations, the harmonic generation spectrum (HGS) obtained from QED is identical to that calculated from Floquet theory, in which the electromagnetic field is treated classically provided that the HG are created when IR absorbed photons are converted to a single high frequency photon.  The key assumption underlying our derivation is that the intensity of the emitted high harmonics is many orders of magnitude smaller than that of the IR laser.}}

\acknowledgments{Prof. Zohar Amitay from the Technion is acknowledged for helpful discussions, insightful comments, and assistance with the HGS calculations using the (t,t') method. This research was partially supported by the Israel Science Foundation (ISF) grant No. 1757/24.}

\bibliographystyle{vancouver}
\bibliography{jpb}

\end{document}